\documentclass[10pt,a4paper,twoside]{article}

\usepackage{a4wide}

\usepackage[utf8]{inputenc} 
\usepackage[TS1,T1]{fontenc}

\usepackage{amssymb, amsmath,subeqnarray}

\usepackage{mathrsfs}

\usepackage{bbold}
\DeclareFontFamily{U}{bbold}{}
\DeclareFontShape{U}{bbold}{m}{n}{<-5.5> bbold5 <5.5-7.5> bbold7 <7.5-> bbold10}{}

\makeatletter

\@addtoreset{equation}{section}
\makeatother

\usepackage{pdfsync}

\usepackage[french,english]{babel}

\title{\textbf{Thermal contact through a two-temperature kinetic Ising chain}}
\author{
M. Bauer\\
 Institut de Physique Théorique de Saclay
 \\  CEA-Saclay \& CNRS
\\ F-91191 Gif-sur-Yvette Cedex, France
\\ \&
\\ Département de Mathématiques et Applications
\\ École Normale Supérieure
\\ PSL Research University
\\ F-75005 Paris, France \\ \vspace{3mm}\\
 F. Cornu\\ Laboratoire de Physique Théorique,  Bât. 210\\
CNRS \& Université Paris-Sud \\
Université Paris-Saclay
\\ F-91405 Orsay Cedex, France
}

\date{\today}

\newcommand{\beq}{\begin{equation}}
\newcommand{\eeq}{\end{equation}}
\newcommand{\bea}{\begin{eqnarray}}
\newcommand{\eea}{\end{eqnarray}}

\newcommand{\ad}{\text{\mdseries a}}

\newcommand{\Amatrix}{\mathbb{A}}

\newcommand{\betae}{\beta_\text{\mdseries e}}
\newcommand{\betao}{\beta_\text{\mdseries o}}

\newcommand{\bra}[1]{\big(#1\vert}
\newcommand{\brat}[1]{\big(#1}

\newcommand{\Bod}{{\cal B}}

\newcommand{\dd}{\text{\mdseries d}}

\newcommand{\ed}{\text{\mdseries e}}
\newcommand{\etaop}{\eta}

\newcommand{\Esp}[1]{\langle#1\rangle}
\newcommand{\Espst}[1]{\langle#1\rangle_\text{\mdseries \mdseries st}}

\newcommand{\fop}{f}

\newcommand{\gammae}{\gamma_\text{\mdseries e}}
\newcommand{\gammao}{\gamma_\text{\mdseries o}}
\newcommand{\gapE}{\Delta E}

\newcommand{\Heat}{Q}
\newcommand{\Heate}{\Heat_\text{\mdseries e}}
\newcommand{\Heato}{\Heat_\text{\mdseries o}}

\newcommand{\Id}{\mathbb{1}}

\newcommand{\ket}[1]{\vert #1 \big)}
\newcommand{\kB}{k_{\scriptscriptstyle B}}

\newcommand{\K}{K}

\newcommand{\lambdab}{\overline{\lambda}}
\newcommand{\lambdae}{\lambda_\text{\mdseries e}}
\newcommand{\lambdao}{\lambda_\text{\mdseries o}}

\newcommand{\Mmatrix}{\mathbb{M}}

\newcommand{\nvec}{\mathbf{n}}
\newcommand{\nvecprime}{\mathbf{n'}}

\newcommand{\nue}{\nu_\text{\mdseries e}}
\newcommand{\nueb}{\overline{\nu}_\text{\mdseries e}}
\newcommand{\nujb}{\overline{\nu}_j}
\newcommand{\nuo}{\nu_\text{\mdseries o}}
\newcommand{\nuob}{\overline{\nu}_\text{\mdseries o}}
\newcommand{\nus}{\nu_\text{\mdseries s}}

\newcommand{\Prob}{P}
\newcommand{\Probst}{P_\text{\mdseries st}}

\newcommand{\QBod}{{\cal QB}}

\newcommand{\svec}{\mathbf{s}}

\newcommand{\Sig}{S}

\newcommand{\Te}{T_\text{\mdseries e}}
\newcommand{\To}{T_\text{\mdseries o}}

\newcommand{\veccolumn}[2]{\begin{pmatrix} #1 \\ #2 \end{pmatrix}}

\newcommand{\Wmatrix}{\mathbb{W}}

\begin{document}
\maketitle

\maketitle

\begin{small}
\begin{abstract}

We consider a model for thermal contact through a diathermal interface between two macroscopic bodies at different temperatures: an Ising spin chain with nearest neighbor interactions is endowed with a Glauber dynamics with  different temperatures and kinetic parameters on alternating sites. The inhomogeneity of the kinetic parameter is a novelty with respect to the model of Ref.\cite{RaczZia1994} and we exhibit its influence upon the  stationary non equilibrium values of the two-spin correlations at any distance. By  mapping to the dynamics of spin domain walls  and using  free fermion techniques, we determine the scaled generating function for the cumulants of the exchanged heat amounts per unit of time in the long time limit.



\end{abstract}
\end{small}

\clearpage

\section{Introduction}

Thermal contact between two macroscopic bodies  initially at different temperatures  corresponds to a situation where the heat transfer between the two bodies is ensured by  a thin diathermal interface. The latter  may be an immaterial interface between two solids or a diathermal wall between two fluids. Over a time window during which the two macroscopic bodies have negligible energy  variations, they  behave as thermostats with constant thermodynamic temperatures, while the interface is a mesoscopic system with traceable configurations. After a long enough time inside the considered time window the interface tends to a stationary non equilibrium state where the  instantaneous heat current which it receives  from a thermostat  has a non-vanishing mean value.  Even if the interface is described by a model  where its degrees of freedom  obey a (deterministic or stochastic)  microscopic dynamics, there is no general framework, such as Gibbs equilibrium ensemble theory, which would allow to determine the probability distribution of the interface configurations and the corresponding mean instantaneous heat current. Therefore it is most valuable to exhibit  solvable  models which would shed some light into the dependence of the heat instantaneous current upon the model parameters and the temperatures of the two thermostats.

Such a solvable model has been introduced by Racz and Zia  in 1994 \cite{RaczZia1994}. They consider a chain of classical spins with periodic boundary conditions and interacting with a nearest-neighbor ferromagnetic (Ising) interaction. They endow it with a Glauber stochastic dynamics with single-spin flips at a time and such that the spins on odd and even lattice sites are flipped by thermostats at two different temperatures.  In our view, the  model  may be seen as a zig-zag shaped chain inside a very thin strip between two half-planes occupied by  two different thermostats, and where the odd  (even) sites are located on the left (right) side of the strip. The stationary two-spin correlations at any distance as well as  higher order spin correlations have  been extensively studied in Refs.\cite{RaczZia1994,SchmittmannSchmuser2002,MobiliaZiaSchmittmann2004,MobiliaSchmittmannZia2005}.

We point out that the latter model indeed satisfies two requirements needed for  a correct description of a thermal contact between two macroscopic bodies during a transient time window where their temperature can be considered as constant. First the contact must be mediated by changes in the  internal energy of the spin interface. Second, if the energies of the macroscopic bodies were kept tracked of, the transition rates for both the interface configurations and these two energies would obey the detailed balance with the microcanonical  equilibrium probability distribution for these variables of the  whole system.  Then, in the infinite time limit the two bodies and the interface would be at the same temperature;  in a time window where the energy  variations  of the macroscopic bodies are negligible,  the transition rates for the interface depend only on the temperatures of the macroscopic bodies and they obey the local detailed balance \cite{Derrida2007,BauerCornu2015}\footnote{In Ref.\cite{Derrida2007} the  terminology "generalized" detailed balance is used.}. 
In the case of a spin interface the two corresponding requirements become: 1) a transition between two spin configurations involves only one thermostat; 2) the transition rate involving a given thermostat obey the detailed balance at the same temperature. For the  present two-temperature Ising chain the first requirement is obviously fulfilled\footnote{We notice that the first requirement is not satisfied by the Ising chain model of Ref.\cite{CornuHilhorst2017}, where two thermostats at different temperatures act on every spin : the latter model does not describe a situation of thermal contact.}, and the
simplest transition rates which fulfill the second requirement  are those chosen by Glauber in the case of an Ising chain in contact with a unique thermostat\cite{Glauber1963}. (We recall that  Glauber looked for single-spin flip dynamics such that in the infinite time limit  the spin chain indeed relaxes to the canonical equilibrium state at the thermostat temperature and then he chose the simplest transition rates.)

\medskip

On the other hand,   the quantities of interest for exchange processes which have been considered over the last three decades, both theoretically and experimentally, are  amounts of microscopically conserved entities (matter, energy, ...), which are exchanged over a very long time (in the case of thermal contact the corresponding quantities are  the heat amounts received by the interface from each thermostat during a fixed long time interval). They  have been focused on because the scaled generating function for their cumulants per unit of time in the long time limit as well as  its Laplace transform, the large deviation function of the corresponding time-integrated current, have been shown to obey  generic symmetry relations, the so-called  fluctuation relations. The latter relations are derived from  properties of the system dynamics and they are a  milestone of 
the stochastic thermodynamics theory (For a review see Ref.\cite{Seifert2012}.)

In this context it is also interesting to have at hand solvable models where the statistics of the time-integrated currents  can be calculated. For instance, such a model has  been exhibited in the case where the heat transfer between two thermostats is ensured by a wire. The energy quanta are represented as particles whose stochastic dynamics is a Symmetric Simple Exclusion Process (SSEP) with adequate boundary  conditions: particles with hard cores hop on the sites of a one-dimensional lattice with the same hopping rate in both directions, but have different in-coming and out-going rates at the two lattice ends in contact with  two particle reservoirs with different chemical potentials. Then all cumulants of the heat coming out of  one thermostat per unit of time can be calculated in the long time limit \cite{DerridaDoucotRoche2004,BodineauDerrida2007} .

In the present paper we consider  a generalized version of the  diathermal interface model of Ref.\cite{RaczZia1994}: the  two Glauber dynamics which flip the spins on sites with odd or even indices respectively   have not only different temperatures but   also different kinetic parameters (see Sec.\ref{secModel}). 
Our aim  is to calculate the scaled generating function of the joint cumulants per unit of time for the heats flowing out of the two thermostats in the long time limit.  

The difference between the kinetic constants is relevant for two macroscopic bodies made with different materials.  Some of the corresponding kinetic effects   have  been  previously  investigated   in our study of  a very simple model for a one-dimensional interface between two half-planes occupied by bodies at two different temperatures : a model of independent two-spin pairs  where the left-side (right-side) spin in each pair is flipped only  by the thermostat on the same side according to a Glauber dynamics. The whole statistics for the spin configurations and the heat amounts exchanged by every pair with the thermostats have been calculated explicitly \cite{CornuBauer2013}.

\medskip
The method which we use to obtain the scaled generating function for the heat cumulants in the present model is the following.
From its definition this  function can be obtained as the largest eigenvalue of the modified Markov matrix which rules the evolution of the joint probability for the system configurations and 
 the heat  amounts  $\Heato$ and $\Heate$ received on each (odd or even) sublattice.  
 
In order to study energy exchanges  more conveniently, instead of considering the spin configurations, we rather formulate the problem in terms of the position configurations for the domain walls, which sit on the dual lattice. In other words we consider the well-known lattice gas representation on the {\it dual\,} lattice (`antiparallel adjacent spin pair' $\leftrightarrow$ `particle' and `parallel adjacent spin pair' $\leftrightarrow$ `hole') where a particle is in fact a domain wall. The mapping of Glauber spin dynamics to the particle dynamics then includes the possibility for the creation and annihilation of adjacent particle pairs, which correspond to the injection or the loss of energy in the chain respectively\footnote{The latter correspondance has been used for instance in Ref.\cite{FaragoPitard2007} for the calculation of the scaled generating function of the energy  injected in an Ising spin ring  through the random  flips of one spin while all other spins evolve according to  a Glauber dynamics at zero temperature which dissipates energy along the ring.}. Our model with two temperatures and two kinetic parameters is mapped to a reaction-diffusion system with two different creation (annihilation) rates as well as two different hopping rates on spatially alternating sites\footnote{In  Ref.\cite{MobiliaSchmittmannZia2005}  the mapping has been used in the reverse sense in order to study the  relaxation towards the stationary state for the reaction-diffusion system from results obtained for the Ising spin chain dynamics with two temperatures but a unique kinetic parameter.}.

The  crucial point is that, in a suitable basis for the representation of the  configurations of domain wall positions,   the Markov matrix for the evolution of the configuration probability involves only products of two operators : the Markov matrix is mapped to   a free fermion Hamiltonian \cite{Schutz2001}.
Moreover the mapping can be readily generalized for the modified Markov matrix which rules the evolution of the joint probability for a configuration of  domain wall positions  and the  amounts of heat $\Heato$ and $\Heate$.
The latter matrix can be diagonalized by using  free fermions techniques: 
 Jordan-Wigner transformation and antiperiodic Fourier transform. Thus we obtain a block diagonal matrix, made of $4\times 4$ blocks, which can be straighforwardly diagonalized by introducing  four pseudo-fermion operators. 
  (In the case of two pseudo-fermion operators and the associated Bogoliubov-like transformation, see for instance Ref.
\cite{GrynbergStinchcombe1995PRL,GrynbergStinchcombe1995PRE}.)
 The largest eigenvalue  of the modified Markov matrix  is obtained by filling up all pseudo-fermion states associated with an eigenvalue with a positive real part.

\medskip

 The paper is organized as follows. The model is defined in Sec.\ref{secModel} and the relaxation time for the mean global magnetization on each sublattice  is calculated. Moreover by solving a hierarchy of equations for the stationary global two-spin correlations at any distance (see Appendix \ref{secTwospincorrelations}), we determine the mean instantaneous energy current which flows from each thermostat into the spin lattice in the stationary state. The mapping to the dynamics of domain walls on the dual lattice and the associated modified matrix is derived in Sec.\ref{SecMapping}. Its eigenvalues are determined in Sec.\ref{SecEigenvalues} by free fermion techniques. The cumulants are obtained in Sec.\ref{SecCumulants} and various physical regimes are discussed. In conclusion we summarize some finite-size effects and their possible cancellation in heat cumulants.

\section{Model}
\label{secModel}

\subsection{Description of a thermalization process}

We consider a one-dimensional lattice with a finite even number  of sites $L=2N$, where each site $j$  is occupied by a classical  spin  $s_j$ ($s_j=\pm 1$, $j=1,\ldots, 2N$) with periodic boundary condition $s_{j+2N}=s_j$.  Spins  interact via the  Ising ferromagnetic nearest-neighbor interaction with coupling $\K>0$: the  energy of a spin configuration $\svec$ is
\beq
\label{defEnergie}
{\cal E}(\svec)=-\K \sum_{j=1}^{2N}s_{j}s_{j+1}.
\eeq
When the  spin  at site $j$ is flipped, the energy variation of the Ising chain is equal to
\beq
\label{varEnergie}
\Delta {\cal E}(s_j\to -s_j)= s_j\frac{s_{j-1}+s_{j+1}}{2} \gapE,
\eeq
with  $\gapE=4\K$. This variation can take the values $+\gapE$, 0, or $-\gapE$,

The model is endowed with a stochastic dynamics where the spin flips at odd (even) sites are due to energy exchanges with a macroscopic body at  temperature $\To$ ($\Te$) in the course of a thermalization process of the two macroscopic bodies. The transition rates must obey  local detailed balance \cite{Derrida2007,BauerCornu2015}: 
the transition rates $w(s_j \to -s_j)$  and  $w(-s_j \to s_j)$  for two reversed flips of the spin at site $j$, while all other spins are kept fixed, must obey the ratio
\beq
\label{LDB}
\frac{w(s_j \to -s_j)}{w(-s_j \to s_j)}=\ed^{-\beta_j \Delta {\cal E}(s_j\to -s_j)},
\eeq
where $\beta_j$ is the inverse temperature of the thermostat acting on site $j$ :  $\beta_j=1/(\kB T_j)$ where $\kB$ is Boltzmann constant and $T_j$ is equal either to $\To$ or $\Te$, depending on the parity of $j$.
As shown by Glauber \cite{Glauber1963} in the case of a unique temperature, the simplest transition rates which obey the constraint \eqref{LDB} read
\beq
\label{deftauxtrans}
w(s_j \to -s_j)=\frac{\nu_j}{2} \left[1-\gamma_j \frac{s_j \left(s_{j-1}+s_{j+1}\right)}{2}\right],
\eeq
while $w(-s_j \to s_j)$ is given by the latter expression where $s_j$ is replaced by $-s_j$.
In \eqref{deftauxtrans} $\gamma_j $ is the thermodynamic parameter at site $j$,
\beq
\label{defgammaa}
\gamma_j =\tanh\left(\frac{\beta_j \gapE}{2}\right)
\eeq
where $\gamma_j=\gammao$ or $\gammae$, depending on the parity of the site index,
and $\nu_j$ is the  kinetic parameter $\nu_j$ at site $j$. The latter is not determined by the local detailed balance; it can be interpreted as the mean frequency at which the macroscopic body tries to flip the spin at site $j$. Therefore we set $\nu_j=\nuo$ or $\nue$, depending on the parity of the site index (note that when the two kinetic parameters are equal the present model coincides with that of Ref.\cite{RaczZia1994}).
Since the time scale is arbitrary, it is convenient to introduce the dimensionless kinetic parameters $\overline{\nu}_\ad$, with  $\text{a}=\text{o}$ or $\text{e}$, defined as
\beq
\label{defDimensionlessNu}
\overline{\nu}_{\text a}= \frac{\nu_{\text a}}{\nuo +\nue}.
\eeq
They satisfy the relation  $\nuob +\nueb=1$ and, apart from the arbitrary time scale, the model has only three independent parameters, $\gammao$, $\gammae$ and $\nuob$.

The probability $\Prob(\svec;t)$ for the system to be in spin configuration $\svec$ at time $t$ evolves according to the master equation
\beq
\label{MarkovSpins}
\frac{\dd \Prob(\svec;t)}{\dd t}=\sum_j^{2N}w(-s_j \to s_j)\Prob(\svec_j;t)- \left(\sum_j^{2N} w(s_j \to -s_j) \right) \Prob(\svec;t)
\eeq
where $\svec_j$ denotes the spin configuration obtained from $\svec$ by changing $s_j$ into $-s_j$.
 The number  of configurations is finite, and the transition rates allow the system to evolve from any configuration to any other one after  a suitable succession of transitions. Therefore there is a unique  stationary solution of the master equation. In the following we focus on the mean values of global quantities and denote $\Esp{\cdots}$ and $\Espst{\cdots}$ the expectation values calculated with the time-dependent probability $\Prob(\svec;t)$ and the stationary probability $\Probst(\svec)$ respectively.

\subsection{Relaxation of the mean global sublattice magnetizations}
\label{secMagnetization}

 The transition rates are invariant under a global flip of the spins, so that
 a configuration and the corresponding one where all spins are flipped have the same probability in the stationary state. As a result all stationary correlations for an odd number of spins vanish identically; in particular $\Espst{s_j}=0$. As a consequence the mean values of the global magnetizations on the two sublattices, $M_\text{\mdseries o}=\sum_{n=1}^{N}s_{2n-1}$ and $M_\text{\mdseries e}=\sum_{n=1}^{N} s_{2n}$ respectively, vanish in the stationary state,
 \beq
 \Espst{M_\text{\mdseries o}}= \Espst{M_\text{\mdseries e}}=0.
 \eeq
  
The relaxation of the mean values of global sublattice magnetizations is readily studied.
As in the case of the  homogeneous spin chain considered by Glauber \cite{Glauber1963}, the evolution equation for the mean value of the spin at site $j$ reads
\beq
\frac{\dd \Esp{s_j}}{\dd t} = - 2 \Esp{s_j w(s_j \to -s_j)}.
\eeq
According to the expression of the transition rates \eqref{deftauxtrans}
\beq
\frac{\dd \Esp{s_j}}{\dd t} = -\nu_j \left[\Esp{s_j} - \gamma_j \frac{\Esp{s_{j-1}}+\Esp{s_{j+1}}}{2} \right].
\eeq
Then the coupled evolutions of the magnetizations on the two sublattices read
\bea
\frac{\dd \Esp{M_\text{\mdseries o}}}{\dd t} &=& - \nuo\left[ \Esp{M_\text{\mdseries o}} - \gamma_\text{\mdseries o} \Esp{M_\text{\mdseries e}}\right]
\\ \nonumber
\frac{\dd \Esp{M_\text{\mdseries e}}}{\dd t} &=& - \nue\left[ \Esp{M_\text{\mdseries e}} - \gamma_\text{\mdseries e} \Esp{M_\text{\mdseries o}}\right].
\eea
From these equations we retrieve that  both mean magnetizations vanish in the stationary state, as predicted by symmetry arguments. The matrix associated with this system of linear equations has two strictly negative eigenvalues  
$\tfrac{1}{2}(\nuo+\nue)\left[-1\pm\sqrt{(\nuob-\nueb)^2+4 \nuob\nueb\gammao\gammae}\right]$, each of which is associated with a couple of  right and left eigenvectors (the eigenvalues are negative because
$(\nuob-\nueb)^2+4 \nuob\nueb\gammao\gammae=1-4 \nuob\nueb(1-\gammao\gammae)<1$).  For generic values of the initial magnetizations $M_\text{\mdseries o}$ and $M_\text{\mdseries e}$, the  inverse relaxation time $1/t_\text{\mdseries rel}$ to their stationary  value is given by  the opposite of the negative  eigenvalue with the smallest modulus, and the relaxation time $t_\text{\mdseries rel}$ reads
\beq
\label{trel}
t_\text{\mdseries rel}=\frac{2}{\nuo+\nue} \left[ 1-\sqrt{(\nuob-\nueb)^2+4 \nuob\nueb\gammao\gammae}\right]^{-1}.
\eeq

\subsection{Mean global heat current in the stationary state}

The mean instantaneous heat current $\Esp{j_k}$ received by the spin chain at site $k$  from the thermostat at temperature $T_k$ is equal to the expectation value of the variation of the chain energy  when the spin $s_k$ is flipped  times the transition rate for the flip. According to the expressions for the energy variation \eqref{varEnergie}
and for  the transition rates \eqref{deftauxtrans}, the mean instantaneous current reads
\beq
\Esp{j_k}=\K\nu_k\left[ -\gamma_k -\gamma_k \Esp{s_{k-1}s_{k+1}} + \Esp{s_{k-1} s_k} +\Esp{s_k s_{k+1}}\right].
\eeq
Therefore the stationary mean value of the global heat current coming from the thermostat acting on spins at even sites, namely
$
J_\text{\mdseries e}= \sum_{n=1}^N j_{2n},
$
is determined as
\beq
\label{ExpressionJe}
\Espst{J_\text{\mdseries e}}=  N \K \nue \left[ -\gammae -\gammae D^\text{oo}_2 +  D^\text{oe}_1 + D^\text{eo}_1\right]
\eeq
with the following definitions : $D^\text{oo}_2$ is the average over the sublattice of odd sites of the  stationary correlation between two spins separated by two sites,
\beq
D^\text{oo}_2=\frac{1}{N}\sum_{n=1}^N \Espst{s_{2n-1} s_{2n+1}},
\eeq 
$D^\text{oe}_1=(1/N)\sum_{n=1}^N \Espst{s_{2n-1} s_{2n}}$ 
and $D^\text{eo}_1$ has an analogous definition. Similarly the stationary mean value of the global heat current coming from the thermostat acting on spins at odd sites,  $J_\text{\mdseries o}= \sum_{n=1}^N j_{2n-1}$, reads
\beq
\label{ExpressionJo}
\Espst{J_\text{\mdseries o}}= N \K \nuo \left[ -\gammao -\gammao D^\text{ee}_2 +  D^\text{oe}_1 + D^\text{eo}_1\right]
\eeq
with $D^\text{ee}_{2}=(1/N)\sum_{n=1}^N \Espst{s_{2n} s_{2n+2}}$.

\medskip
The values of the stationary global  two-spin correlations $D^\text{oo}_2$, $D^\text{ee}_2$, $D^\text{oe}_1$ and $D^\text{eo}_1$ can be determined from a hierarchy of equations for similar quantities with two spins at any distance on the lattice. Details are given in Appendix \ref{secTwospincorrelations} with the results
 \eqref{relDooDee}, \eqref{relDoeDeo}, \eqref{valueDee},  and \eqref{valueDoe} for any distance between spins. From the latter results we get
\beq
\label{valueDee2}
D^\text{ee}_2=\frac{\gamma  \eta_-}{\gammao} \, \frac{1 +\eta_-^{N- 2} }{1+\eta_-^N} 
\eeq
with
\beq
\label{defgamma}
\gamma= \nuob \gammao + \nueb \gammae,
\eeq
where the dimensionless kinetic parameters have been defined in \eqref{defDimensionlessNu}, and $\eta_-$, with $0< \eta_- <1$, is defined by
\beq
\sqrt{\eta_-}=\frac{1-\sqrt{1-\gammao\gammae}}{\sqrt{\gammao\gammae}}.
\eeq
Moreover $D^\text{oo}_2=(\gammao/\gammae) D^\text{ee}_2$, while
\beq
\label{valueDoe1}
D^\text{oe}_1=D^\text{eo}_1=\frac{\gamma  \sqrt{\eta_-}}{\sqrt{\gammao\gammae}} \,\frac{1 +\eta_-^{N-1}}{1+\eta_-^N}.
\eeq
We notice that the dependence upon the kinetic parameters $\nuo$ and $\nue$ occurs only through the parameter $\gamma$ defined in \eqref{defgamma}.

Eventually the stationary mean value of the global heat current received on even sites can be calculated from  expression \eqref{ExpressionJe} and  relation \eqref{RelRoots};
we get  
\beq
\label{valueJe}
\Espst{J_\text{\mdseries e}}= N \K \frac{\nuo \nue}{\nuo+\nue} (\gammao-\gammae).
\eeq
Similarly the stationary mean value  of the  global heat current received on odd sites can be obtained from the expression \eqref{ExpressionJo} ; it  proves to be opposite to that on even sites, 
$\Espst{J_\text{\mdseries o}}=-\Espst{J_\text{\mdseries e}}$, as it should in the stationary state where the mean  energy of the chain is constant.

We point out the following remarkable property : though $D^\text{ee}_2$, $D^\text{oo}_2$, $D^\text{oe}_1$ and $D^\text{eo}_1$  involve finite size corrections (see \eqref{valueDee2} and \eqref{valueDoe1}), these corrections cancel one another in the value of the mean global current $\Espst{J_\text{\mdseries e}}$ : $\Espst{J_\text{\mdseries e}}$ is exactly proportional to the size $L=2N$ of the ring. Moreover it happens to be equal to $L$ times the mean current received by a spin in the independent pair model of Ref.\cite{CornuBauer2013}.

\section{Mapping to a reaction-diffusion system with pair creation-annihilation}
\label{SecMapping}

To prepare the study of the heat amounts exchanged with the thermostats we consider a mapping to another model for which the evolution operator is quadratic. 

\subsection{Domain wall system}

When two spins on neighboring sites are antiparallel, one may consider that there is a domain wall between them, whereas there is no domain wall when they are parallel. The domain walls sit on the edges of the initial lattice. Labeling each edge by its mid-point, one gets another lattice which we call the dual lattice in what follows. The edge $(j-1,j)$ and the corresponding site on the dual lattice are labeled by $j$.
 If $s_{j-1}$ and $s_j$ are antiparallel, $s_{j-1}s_j=-1$, then the occupation number by a domain wall at site $j$ on the dual lattice is $n_j=1$, whereas if $s_{j-1}$ and $s_j$ are parallel $n_j=0$. Thus the correspondance reads
\beq
\label{CorrespondenceWallSpin}
n_j=\frac{1-s_{j-1}s_{j}}{2}.
\eeq
On a ring the number of domain walls is even and $\sum_{j=1}^L n_j$ is even.

As a result a spin configuration can be characterized either by the set $\svec=\{s_1,\cdots s_L\}$ of spin configurations or by the knowledge of the value of $s_1$ and the set of the positions of the domains walls, 
namely the set of occupations numbers $\nvec=\{n_1,\cdots,n_L\}$. 
The energy of the system can be expressed solely in terms of   domain walls as
\beq
{\cal E}(\nvec)= - 2N \K+2 \K \sum_{j=1}^{2N} n_j.
\eeq

\subsection{Quantum mechanics notations}

In the following we use the quantum mechanics notations, as commonly done in the literature. Then a column vector is denoted as  a  ``ket'', $\ket{\ldots}$ and a row vector is denoted as a ``bra'' $\bra{\ldots}$.
The configuration of occupation numbers by domain walls ${\nvec}=\{n_1, \cdots, n_{L}\}$ is represented as a tensor product
\beq
\label{tensorialprod}
 \ket{\nvec}=\otimes_{j=1}^{L} \ket{n_j},
\eeq
where $\ket{n_j}$ is a two-component column vector.
The convention used for kets associated to vacant and occupied states is 
\beq
\label{convention-vacant}
\veccolumn{1}{0}_j=\ket{n_j=0}\quad\textrm{and}\quad \veccolumn{0}{1}_j=\ket{n_j=1}.
\eeq
This convention is the standard choice of basis in the condensed matter literature on quantum spin chains.
With the representation \eqref{tensorialprod}-\eqref{convention-vacant} the row-column product $\brat{\nvecprime}\ket{\nvec}$ takes the form
$
\brat{\nvecprime}\ket{\nvec}=\prod_{j=1}^{L} \delta_{n'_j,n_j}
$.
Therefore the probability of the domain wall configuration $\nvec$ at time $t$, $\Prob(\nvec;t)$,
can  be represented as a row-column (scalar) product 
$
\Prob(\nvec;t)= \brat{\nvec} \ket{\Prob_t}
$,
where $\ket{\Prob_t}$ is  the column vector defined as
\beq
\label{defketProb}
\ket{\Prob_t}=\sum_{\nvec} \Prob(\nvec;t) \ket{\nvec}.
\eeq

With the latter definitions the master equation for the stochastic evolution of the probability $\Prob(\nvec;t)$, which takes the generic form written in  \eqref{MarkovSpins} in the case of $\Prob(\svec;t)$, can be represented as the evolution of the column vector $\ket{\Prob_t}$ under the Markov matrix $\Mmatrix$
\beq
\frac{\dd \ket{\Prob_t}}{\dd t}=\Mmatrix \ket{\Prob_t}
\eeq
with 
\bea
\bra{\nvec'} \Mmatrix \ket{\nvec}&=& w(\nvec \to \nvec')
 \quad\textrm{if}\quad \nvec'\neq \nvec
\\ \nonumber
\bra{\nvec} \Mmatrix \ket{\nvec}&=&-\sum_{\nvec'\neq\nvec} w(\nvec \to \nvec'),
\eea
where $w(\nvec \to \nvec')$ denotes the transition rate from  configuration $\nvec$ to  configuration $\nvec'$.

\subsection{Markov matrix for the model}

For the present model of domain walls the  matrix elements of $\Mmatrix$ can be expressed in terms of Pauli matrices. Indeed the  operator for the occupation number at site $j$ reads 
\beq
\label{widehatn}
\widehat{n}_j=\tfrac{1}{2} \left( \Id_j-\sigma^z_j\right)
\eeq
where $\Id_j$ denotes the identity $2\times 2$ matrix at site $j$, and
{\small $\sigma^z_j=\begin{pmatrix}
   1 & 0 \\
   0 & -1
\end{pmatrix}_j$};
 the operator which changes the occupation number  at site $j$  is 
{\small $
\sigma^x_j=
\begin{pmatrix}
   0 & 1 \\
   1 & 0
\end{pmatrix}_j
$}.
By inspection of the  transition rates for the spin configurations $w(\svec \to \svec')$ given by \eqref{deftauxtrans}, every  transition rate for the occupation numbers by domain walls, $w(\nvec \to \nvec')$, can be written as a matrix element $\bra{\nvec'} \Wmatrix \ket{\nvec}$, where 

- for a hop of a domain wall from site $j$ to site $j+1$
\beq
\Wmatrix =\frac{\nu_j}{2}\sigma^x_j \sigma^x_{j+1} \widehat{n}_j \left(\Id_j-\widehat{n}_{j+1}\right)
\eeq

- for a hop of a domain wall  from site $j+1$ to site $j$
\beq
\Wmatrix =\frac{\nu_j}{2}\sigma^x_j \sigma^x_{j+1} \left( \Id_j-\widehat{n}_j\right) \widehat{n}_{j+1}
\eeq

- for the annihilation of two domain walls at sites $j$ and $j+1$ 
\beq
\Wmatrix = \frac{\nu_j}{2}(1+\gamma_j)\sigma^x_j \sigma^x_{j+1} \widehat{n}_j \widehat{n}_{j+1}
\eeq

- for the creation of two domain walls at sites $j$ and $j+1$ 
\beq
\Wmatrix =\frac{\nu_j}{2}(1-\gamma_j)\sigma^x_j \sigma^x_{j+1}  \left( \Id_j-\widehat{n}_j\right)  \left( \Id_j-\widehat{n}_{j+1}\right).
\eeq
The latter expressions can be written in a more compact form by using the 
 spin-$\tfrac{1}{2}$ ladder operators 
{\small $
\sigma^+=
\begin{pmatrix}
   0 & 1 \\
   0 & 0
\end{pmatrix}_j
$} and 
{\small $
\sigma^-=
\begin{pmatrix}
   0 & 0 \\
   1 & 0
\end{pmatrix}_j
$}. They are such that $\sigma^x\widehat{n}=\sigma^+$ and $\sigma^x(\Id-\widehat{n})=\sigma^-$.
With the convention \eqref{convention-vacant} $\sigma^+_j$ annihilate a domain wall at site $j$, while $\sigma^-_j$ creates a domain wall at site $j$.

Eventually the Markov matrix $\Mmatrix$  derived from the master equation for the evolution  of the probability of  spin configurations \eqref{MarkovSpins} reads
\bea
&&\Mmatrix =\frac{\nu_1+\nu_2}{2}
\\ \nonumber
&& \times
 \left[ -(1-\gamma) N \Id
 -\gamma\sum_{j=1}^{2N} \widehat{n}_j 
+\sum_{j=1}^{2N} \nujb\left[ \sigma^+_j \sigma^-_{j+1}  + \sigma^-_j \sigma^+_{j+1}+
 (1+\gamma_j)\sigma^+_j \sigma^+_{j+1} +(1-\gamma_j)\sigma^-_j \sigma^-_{j+1} \right]\right],
\eea
where $\Id$ denotes the identity $2N \times 2N$ matrix and $\gamma$ has been defined in \eqref{defgamma}. The advantage of the domain wall representation with respect to the spin representation is that the Markov matrix is quadratic in terms of operators acting on different sites instead of involving three operators acting on different sites (for the latter case see for instance Ref.\cite{Felderhof1971-A,Felderhof1971-B}).

\section{Eigenvalues of the modified Markov matrix}
\label{SecEigenvalues}
 
\subsection{Modified Markov matrix}
\label{CumulantsMaximumEigenvalue}

We are interested in the joint cumulants per unit of time for  the heat amounts $\Heato$ and $\Heate$ which are received by the chain from the thermostat acting on spins at odd  and even sites during a time $t$ in the long time limit. The corresponding scaled generating function  is
\beq
\label{defgL}
g_{2N}(\lambdao,\lambdae;t)= \lim_{t\to\infty} \frac{1}{t}  \ln\Esp{ \ed^{\lambdao\Heato+\lambdae\Heate}},
\eeq
where $\lambdao$ and $\lambdae$ are real parameters.
In fact  an evolution equation can be written for the probability 
$\Prob(\nvec, \Heato,\Heate;t)$ for the system to be in configuration $\nvec$ at time $t$ and to have received  heat amounts $\Heato$  and $\Heate$ between times $0$ and $t$. Therefore 
the expectation value in the definition \eqref{defgL} can be expressed  in terms of the discrete Laplace transform of   
$\Prob(\nvec, \Heato,\Heate;t)$, and then $g_{2N}(\lambdao,\lambdae;t)$ reads
\beq
\label{defgLBis}
g_{2N}(\lambdae,\lambdao;t)= \lim_{t\to\infty} \frac{1}{t}  \ln\sum_{\nvec} \widehat{\Prob}(\nvec, \lambdao,\lambdae;t)
\eeq
with
\beq
\widehat{\Prob}(\nvec, \lambdao,\lambdae;t)= \sum_{\{\Heato,\Heate\}} \ed^{\lambdao\Heato+\lambdae\Heate}\Prob(\nvec, \Heato,\Heate;t).
\eeq

 By inspection of the transition rates \eqref{deftauxtrans} and according to the correspondence \eqref{CorrespondenceWallSpin},
 when the spin at site $j$ is flipped under the action of the thermostat at temperature $T_j$, the variation of 
$\Heat_j$ is equal to $+\gapE$ if a pair of domain walls  is created at sites $j$ and $j+1$ , $-\gapE$ if a  pair of domain walls is annihilated at these sites, $0$ if a domain wall jumps either from $j$ to $j+1$ or from $j+1$ to $j$.
As a consequence, with a definition for $ \ket{\widehat{\Prob}_t (\lambdao,\lambdae)}$ analogous to that for $\ket{\Prob_t}$ given in \eqref{defketProb}, namely $\brat{\nvec}\ket{\widehat{\Prob}_t (\lambdao,\lambdae)}=\widehat{\Prob}(\nvec, \lambdao,\lambdae;t)$, we get the evolution equation
\beq
\label{EvLPProb}
\frac{\dd \ket{\widehat{\Prob}_t (\lambdao,\lambdae)}}{\dd t}= \widehat{\Mmatrix}(\lambdao,\lambdae) \ket{\widehat{\Prob}_t(\lambdao,\lambdae)},
\eeq
where  the so-called modified Markov matrix $\widehat{\Mmatrix}(\lambdao,\lambdae)$ reads
\beq
\label{defMmodified}
\frac{2}{\nu_1+\nu_2} \widehat{\Mmatrix}(\lambdao,\lambdae) =-(1-\gamma) N \Id -\gamma\sum_{j=1}^{2N} \widehat{n}_j 
+\sum_{j=1}^{2N} \nujb\left[ \sigma^+_j \sigma^-_{j+1}  + \sigma^-_j \sigma^+_{j+1}+
b_j\sigma^+_j \sigma^+_{j+1} +c_j \sigma^-_j\sigma^-_{j+1} \right].
\eeq
The coefficient  $b_j$  ($c_j$) is  equal either to $b_\text{\mdseries o}$ or $b_\text{\mdseries e}$ ($c_\text{\mdseries o}$ or $c_\text{\mdseries e}$) according to the parity of $j$;   with the notation $\text{a}=\text{o}$ or $\text{e}$
\beq
\label{defbj}
b_\ad=(1+\gamma_\ad)\,\ed^{-\lambda_\ad \gapE}
\eeq
and
\beq
\label{defcj}
c_\ad=(1-\gamma_\ad)\,\ed^{\lambda_\ad\gapE}.
\eeq
According to the evolution equation \eqref{EvLPProb} the Laplace transform   $\widehat{\Prob}(\nvec, \lambdao,\lambdae;t)$ is equal to \\ $\bra{\nvec} \exp[\widehat{\Mmatrix}(\lambdao,\lambdae) t] \ket{\widehat{\Prob}_{t=0} (\lambdao,\lambdae)}$. 
Thus the scaled generating function $g_{2N}(\lambdao,\lambdae;t)$ given by \eqref{defgLBis} is equal to the largest eigenvalue of the matrix $\widehat{\Mmatrix}(\lambdao,\lambdae) $ which rules the evolution of $\ket{\widehat{\Prob}_t(\lambdao,\lambdae)}$.

\subsection{Jordan-Wigner transformation}

In order to find the eigenvalues of the modified Markov  matrix  $\widehat{\Mmatrix}(\lambdao,\lambdae)$ given by \eqref{defMmodified} we take advantage of its structure analogous to a free fermion Hamiltonian and we introduce the following Jordan-Wigner transformation \cite{JordanWigner1928}
\beq
\fop^\dag_j=\left(\prod_{k=1}^{j-1}\sigma^z_k \right)\sigma^-_j
\quad\textrm{and}\quad
\fop_j=\left(\prod_{k=1}^{j-1}\sigma^z_k \right)\sigma^+_j.
\eeq
The operator $\fop_j^\dag$ is indeed the adjoint  of $\fop_j$, because $(\sigma^z)^\dag=\sigma^z$ and $(\sigma^-)^\dag=\sigma^+$.
Operators $\sigma$ acting on different sites commute, whereas $\sigma^z_j$ anticommutes with $\sigma^+_j$ and $\sigma^-_j$; moreover $(\sigma^+_j)^2=0$ and $(\sigma^-_j)^2=0$. Therefore the operators $\fop_j$ and $\fop^\dag_j$ obey the fermionic anticommutation relations
\beq
\{\fop_j,\fop_{j'}\}=0 \quad \{\fop^\dag_j,\fop^\dag_{j'}\}=0 \quad \{\fop_j,\fop^\dag_{j'}\}=\delta_{j,j'}.
\eeq
The occupation number of site $j$ by a domain wall, given by  \eqref{widehatn},   also reads $\widehat{n}_j=\sigma^-_j\sigma^+_j=\fop^\dag_j \fop_j$. The expression \eqref{defMmodified} of the modified matrix $\widehat{\Mmatrix}(\lambdao,\lambdae)$ is rewritten in terms of fermionic operators as
\bea
\nonumber
\label{widehatM}
\frac{2}{\nu_1+\nu_2} \widehat{\Mmatrix}(\lambdao,\lambdae) 
&=&-(1-\gamma) N \Id -\gamma\sum_{j=1}^{2N}\fop^\dag_j\fop_j
+\sum_{j=1}^{2N-1} \nujb\left[\fop^\dag_j \fop_{j+1} - \fop_j\fop^\dag_{j+1} +c_j \fop^\dag_j\fop^\dag_{j+1} -b_j \fop_j\fop_{j+1} \right] 
\\
&-& \nueb (-1)^{{\cal N}_\text{\mdseries f}}\left[ \fop^\dag_{2N} \fop_1 - \fop_{2N}\fop^\dag_1 +c_\text{\mdseries e} \fop^\dag_{2N}\fop^\dag_1-b_\text{\mdseries e} \fop_{2N}\fop_1\right]
\eea
where ${\cal N}_\text{\mdseries f}=\sum_{j=1}^{2N} \fop^\dag_j \fop_j$ is the total number of fermions.

Since the spin system is on a ring, there can be only an even number of domain walls in the system. As noticed above, the operator for the occupation number by a domain wall $\widehat{n}_j$ coincides with the number of fermions at site $j$, $\fop^\dag_j \fop_j$. Therefore we have to consider  the restriction  $\widehat{\Mmatrix}_+(\lambdao,\lambdae) $ of  $\widehat{\Mmatrix}(\lambdao,\lambdae)$  to the sector with an even number of fermions.
According to \eqref{widehatM} the expression of this rectriction   is invariant by translation along the ring if the fermionic operators are chosen to satisfy the antiperiodic boundary conditions
\beq
\label{AntiperiodicC}
\fop_{2N+1}=-\fop_1 \quad\textrm{and}\quad \fop^\dag_{2N+1}=-\fop^\dag_1.
\eeq
Then the restriction reads
\beq
\label{Mplus}
\frac{2}{\nu_1+\nu_2} \widehat{\Mmatrix}_+(\lambdao,\lambdae) 
=-(1-\gamma) N \Id -\gamma\sum_{j=1}^{2N}\fop^\dag_j\fop_j
+\sum_{j=1}^{2N} \nujb\left[\fop^\dag_j \fop_{j+1} - \fop_j\fop^\dag_{j+1} +c_j \fop^\dag_j\fop^\dag_{j+1} -b_j \fop_j\fop_{j+1} \right].
\eeq

\clearpage
\subsection{Antiperiodic Fourier transform}

The next step to the diagonalization is to rewrite the fermionic operators as antiperiodic Fourier transforms which satisfy the antiperiodic boundary conditions \eqref{AntiperiodicC}
The  wave numbers are of the form  $q=(2k+1)\pi/(2N)$ and we work with a complete family of representatives in the set 
\beq
\label{defBod2N}
\Bod(2N)=\{q=(2k+1) \frac{\pi}{2N}, k=-N,-N+1, \cdots, -1,0,1, \cdots N-1 \},
\eeq
namely
\beq
\Bod(2N)=\{-\pi +\frac{\pi}{2N},  -\pi +\frac{3\pi}{2N}, \cdots -\frac{\pi}{2N}, \frac{\pi}{2N}, \cdots \pi-\frac{\pi}{2N}  \}.
\eeq
The operator $\fop_j$ can be written as the antiperiodic Fourier transform
\beq
\label{defAntiFT}
\fop_j=\frac{1}{\sqrt{2N}} \sum_{q\in \Bod(2N)} \ed^{\imath qj}  \etaop_q
\eeq
in terms of the wave fermions
\beq
\label{defAntiFTInv}
\eta_q =\frac{1}{\sqrt{2N}} \sum_{j=1}^{2N} \ed^{-\imath qj} f_j.
\eeq
Going from  \eqref{defAntiFT}  to \eqref{defAntiFTInv} relies on the identity
\beq
\label{SumGeom}
\sum_{j=1}^{2N} \ed^{\imath(q-q')j}=2N \,\mathbb{1}_{q-q'\equiv 0 (2\pi)},
\eeq
where $\mathbb{1}_{q-q'\equiv 0 (2\pi)}=1$ if $q-q'$ is equal to $0$ modulo $2\pi$ and $\mathbb{1}_{q-q'\equiv 0 (2\pi)}=0$ otherwise.
All $q'$s in $\Bod(2N)$ satisfy $\ed^{\imath q 2N}=-1$, and subsequently $\fop_j$  does obey the antiperiodic boundary conditions \eqref{AntiperiodicC}.
The adjoint operator  $\fop^\dag_j$ reads
\beq
\fop^\dag_j=\frac{1}{\sqrt{2N}} \sum_{q\in \Bod(2N)} \ed^{-\imath qj}  \etaop^\dag_q.
\eeq

These representations are inserted in the expression \eqref{Mplus}. In the term  $\sum_{j=1}^{2N}\fop^\dag_j\fop_j$ there occurs  a summation over all sites of the ring and one uses the identity \eqref{SumGeom}.
 In the other summations  one has to distinguish the two sublattices ; for instance one has to consider the sum $\sum_{n=1}^N f^\dag_{2n} f^\dag_{2n+1}$ and then one uses the identity
\beq
\sum_{n=1}^{N} \ed^{\imath(q+q')2n}=N \,\mathbb{1}_{2(q+q')\equiv 0 (2\pi)}=N \,\mathbb{1}_{q+q'\equiv 0(\pi)}.
\eeq
According to  definition \eqref{defBod2N}, the set $\Bod(2N)$ does not contain  the value $0$ and the solution of $q+q'=0$ corresponds to two distinct values $q$ and $-q$.

In order to simplify the following discussion, we assume from now on that $N$ \textit{is  even}. Then  $\Bod(2N)$ does not contain $\pi/2$ and all values $q$ and $\pi- q$, are also distinct. 
After a symmetrization over the values $q$ and $\pi-q$  the matrix  $\widehat{\Mmatrix}_+(\lambdao,\lambdae)$ appears as a sum of contributions each of which involves only the  operators associated with the wave numbers $q$, $\pi-q$, $-q$ and $-(q-\pi)$. Let us introduce 
the first quadrant in the set $\Bod$ defined as
\beq
\label{defQBod}
\QBod(2N)=\{q=(2k+1) \frac{\pi}{2N}, k=0,1, \cdots (N/2) -1 \} =\{ \frac{\pi}{2N}, \frac{3\pi}{2N}\cdots \frac{\pi}{2}-\frac{\pi}{2N}\}.
\eeq
Then the expression \eqref{Mplus} for
$\widehat{\Mmatrix}_+(\lambdao,\lambdae)$ can be rewritten as
\beq
\label{widehatMBis}
\frac{2}{\nu_1+\nu_2} \widehat{\Mmatrix}_+(\lambdao,\lambdae) 
=-N \Id  +\sum_{q\in \QBod} \left[V^\dag_q \right]^{\scriptscriptstyle T} \Amatrix_q(\lambdao,\lambdae)\, V_q,
\eeq
where $V_q$ is the column vector
\beq
\label{defVq}
V_q=
\begin{pmatrix}
\etaop_q \\
\etaop_{q-\pi} \\
\etaop^\dag_{-q} \\
\etaop^\dag_{\pi-q}
\end{pmatrix},
\eeq
$\left[V^\dag_q \right]^{\scriptscriptstyle T}$ denotes the transposed row vector corresponding to the column vector $V^\dag_q$ built with the adjoints of the components of $V_q$, and
\beq
\Amatrix_q (\lambdao,\lambdae) =
\begin{pmatrix}
 -\gamma + \cos q & \imath a_q \sin q & \imath c'_q \sin q & c_q \cos q \\
 - \imath a_q \sin q  &   -\gamma - \cos q & -c_q \cos q & -\imath c'_q \sin q \\
- \imath b'_q \sin q  & - b_q \cos q & \gamma - \cos q & -\imath  a_q \sin q \\
b_q \cos q  & \imath b'_q \sin q & \imath a_q \sin q & \gamma + \cos q
\end{pmatrix},
\eeq
with 
\bea
a_q &=& \nuob-\nueb
\\ \nonumber
b_q&=& \nuob b_\text{\mdseries o} -\nueb b_\text{\mdseries e}
\\ \nonumber
b'_q&=& \nuob b_\text{\mdseries o} +\nueb b_\text{\mdseries e} 
\\ \nonumber
c_q&=& \nuob c_\text{\mdseries o} -\nueb c_\text{\mdseries e}
\\ \nonumber
c'_q&=& \nuob c_\text{\mdseries o} +\nueb c_\text{\mdseries e},
\eea
where the $b_\ad$'s  and the $c_\ad$'s are defined in \eqref{defbj} and \eqref{defcj}. 

\subsection{Diagonalization of $\Amatrix_q(\lambdao,\lambdae) $}

The characteristic polynomial of  $\Amatrix_q$, $\text{det}\left(\Amatrix_q-\alpha \Id_q\right)$,  proves to be a second order polynomial  in $\alpha^2$, with a constant which is a squared quantity,
\beq
\label{CharacteristicPol}
\text{det}\left(\Amatrix_q-\alpha \Id_q\right)= \alpha^4 -2 D \alpha^2 +F^2.
\eeq
Moreover both coefficients $D$ and $F^2$ depend on the parameters $\lambdao$ and $\lambdae$ only through the difference
\beq
\label{deflambdab}
\lambdab=\left(\lambdae-\lambdao\right)\gapE.
\eeq
They read
\beq
D=1 + (\nuob-\nueb)^2 + \nuob\nueb \left[4 \gammao\gammae\cos^2q  + (1-2\cos^2q) \theta(\lambdab) \right]
\eeq
and 
\beq
F=\nuob\nueb \left[ 4(1- \gammao\gammae \cos^2 q) + \theta(\lambdab)\right],
\eeq
where the function $\theta(\lambdab)$  vanishes when $\lambdab$ is set to zero :
\beq
\label{deftheta}
\theta(\lambdab)=2\left[ (1-\gammao\gammae) (\cosh \lambdab -1) + (\gammao-\gammae)\sinh\lambdab \right].
\eeq
The coefficient $F$ can be rewritten as 
\beq
F=\nuob\nueb \left[ 2 + 2 \gammao\gammae(1-2 \cos^2 q) + (1-\gammao\gammae) \cosh \lambdab + (\gammao-\gammae)\sinh\lambdab\right],
\eeq
and the property $(1-\gammao\gammae) > \vert \gammao-\gammae \vert$ for $\gamma_\text{\mdseries o}<1$ and $\gamma_\text{\mdseries e}<1$ ensures that $F>0$.
The squared  roots of the characteristic polynomial are
\beq
\label{valuealpha2}
\alpha^2=D \pm \sqrt{D^2-F^2} = \left(\sqrt{\frac{D+F}{2}} \pm \sqrt{\frac{D-F}{2}} \right)^2.
\eeq
where $\sqrt{\cdots}$ denotes a possibly complex square root. Let us introduce the notations
$R_\pm(q,\lambdab)= \tfrac{1}{2}(D\pm F)$, namely
\beq
\label{defRplus}
R_+(q,\lambdab)= 1+\nuob\nueb \theta(\lambdab) \sin^2q
\eeq
\beq
\label{defRminus}
R_-(q,\lambda)=(\nuob-\nueb)^2+\nuob\nueb \left[4 \gammao\gammae  -\theta(\lambdab)\right]\cos^2 q.
\eeq
We notice that, since  $F$ is positive, $R_+> R_-$.

Note that  $1+\nuob\nueb \theta(\lambdab)>0$ because of the definition \eqref{deftheta} of $\theta(\lambdab)$ and the identities
 $1-2\nuob\nueb (1-\gammao\gammae) >0$ and $(1-\gammao\gammae)> \vert \gammao-\gammae \vert$
  for $\gammao<1$ and $\gammae<1$.  According to \eqref{defRplus}, $R_+$ can be rewritten as 
$
R_+=1-\sin^2 q + 
\left(1+\nuob\nueb \theta(\lambdab)\right)\sin^2q
$ and we conclude that 
$R_+>0$. 

On the other hand, by virtue of  definitions \eqref{defgammaa} and \eqref{deftheta},
 \beq
 \label{relthetacosh}
4 \gammao\gammae  -\theta(\lambdab)=2\frac{\cosh\left((\betae+\betao) \gapE/2\right)-\cosh\left(\lambdab-(\betae-\betao) \gapE/2\right)}{\cosh\left(\betae \gapE/2\right)\cosh\left(\betao \gapE/2\right)},
\eeq
so that  $4 \gammao\gammae  -\theta(\lambdab)>0$ only  if  $-\betao \gapE<
\lambdab<\betae\gapE$, and we infer from \eqref{defRminus} that $R_-$ can take both signs. 

Eventually the four eigenvalues of $\Amatrix_q(\lambdao,\lambdae)$ are
\beq
\label{defalphan}
\alpha_1= \sqrt{R_+} + \sqrt{R_-} \qquad \alpha_2= \sqrt{R_+} - \sqrt{R_-}\qquad \alpha_3=-\alpha_2 \qquad \alpha_4=-\alpha_1,
\eeq
In these expressions $\sqrt{R_+}$ denotes the usual positive square root of the positive number $R_+$, whereas $\sqrt{R_-}$ is either real positive or purely imaginary depending on the sign of $R_-$, i.e. on the values of $\lambdab$ and $q$.
 As noticed previously  $R_+> R_-$, and  in the case where $\sqrt{R_-}$ is real, all $\alpha_k$'s, with $k=1,\ldots,4$, are real and $\alpha_1>\alpha_2>0> \alpha_3 > \alpha_4$.

\subsection{Largest eigenvalue of the modified matrix $\widehat{\Mmatrix}_+(\lambdao,\lambdae)$}

For the sake of conciseness we omit all dependences upon $\lambdao$ and $\lambdae$ in the present section.
We denote by $\mathbb{D}_q$ the diagonal matrix built with the eigenvalues $\alpha_1(q),\ldots,\alpha_4(q)$ of the matrix $\Amatrix_q(\lambdao,\lambdae)$.  The matrix $\Amatrix_q$ reads
\beq
\label{relAD}
\Amatrix_q= \mathbb{P}_q \mathbb{D}_q \mathbb{P}_q^{-1},
\eeq
where the $k^\text{th}$ column of $\mathbb{P}_q$ is made with the components of a (column) right eigenvector of 
$\Amatrix_q$ associated with the eigenvalue $\alpha_k(q)$, and $\mathbb{P}_q^{-1}$ is the inverse matrix of  $\mathbb{P}_q$. Let $\xi_k(q)$  denote the $k^\text{th}$ component of the column vector $\mathbb{P}_q^{-1} V_q$, while  $\xi^\star_k(q)$ denotes the 
$k^\text{th}$ component of the row vector $\left[V^\dag_q \right]^{\scriptscriptstyle T} \mathbb{P}_q$, where $V_q$ and $\left[V^\dag_q \right]^{\scriptscriptstyle T} $  are defined in \eqref{defVq}. With these definitions, the relation \eqref{relAD} implies that
\beq
\label{Sumxi}
\left[V^\dag_q \right]^{\scriptscriptstyle T} \Amatrix_q V_q=\sum_{k=1}^4 \alpha_k(q) \, \xi^\star_k(q) \,\xi_k(q).
\eeq
The operators $\xi_k$ and $\xi^\star_k$ obey the anticommutation rules $\{\xi_k, \xi_{k'}\}=0$, $\{\xi^\star_k, \xi^\star_{k'}\}=0$ and $\{\xi_k, \xi^\star_{k'}\}=\delta_{k,k'}$. However, since $\Amatrix_q$ is not hermitian (for the usual scalar product), 
$\mathbb{P}_q$ is not unitary and the operator $\xi^\star_k$ is not  the adjoint of $\xi_k$.

Nevertheless the anticommutation rules are enough to ensure that  the spectrum  of the operator $\xi^\star_k \xi_k$ is the set $\{0,1\}$. Then, according to the expressions \eqref{defalphan} of the $\alpha_k$'s, the  value of the right-hand side of \eqref{Sumxi} with the largest real part  is equal to the sum of two eigenvalues and proves to be real positive,
\beq
\alpha_1 + \alpha_2=2 \sqrt{R_+(q)}>0.
\eeq
Eventually, according to \eqref{widehatMBis},   the largest eigenvalue of $ \widehat{\Mmatrix}_+(\lambdao,\lambdae)$ is
\beq
\label{LargestEigenvalue}
\frac{\nuo+\nue}{2}
\left[-N   + 2\sum_{q\in \QBod} \sqrt{R_+(q)}\right],
\eeq
where $R_+(q)$ is given by \eqref{defRplus}.

We notice that if $\lambdao=\lambdae=0$, the modified Markov matrix $\widehat{\Mmatrix}(\lambdao,\lambdae)$  becomes the usual Markov matrix for the evolution of $\Prob(\nvec;t)$; since $\theta(0)=0$, we retrieve that the largest eigenvalue of the Markov matrix is $0$. Moreover the eigenvalue closest to 0 is obtained by setting $ \xi^\star_3(q)\xi_3(q)$ equal to 1 for $q=0$, and we retrieve the value \eqref{trel} for the inverse relaxation time of the magnetizations on the two sublattices.

\clearpage
\section{Cumulants of heat amounts  per unit of time in the long time limit}
\label{SecCumulants}

\subsection{Scaled generating function for joint cumulants}

According to the remark at the end of section \ref{CumulantsMaximumEigenvalue}, the scaled generating function for the joint cumulants of $\Heato$ and $\Heate$ coincides with the largest eigenvalue of $ \widehat{\Mmatrix}_+(\lambdao,\lambdae)$. By virtue of \eqref{LargestEigenvalue} it reads
\beq
\label{ExpgL}
g_{2N}(\lambdao,\lambdae)= \frac{\nuo+\nue}{2}\left[-N + 2 \sum_{q\in \QBod} \sqrt{1+\nuob\nueb \theta(\lambdab) \sin^2q}\right],
\eeq
where $\QBod(2N)$ is defined in \eqref{defQBod} and $\theta(\lambdab)$ is given in \eqref{deftheta}.
The joint cumulants per unit of time in the long time limit are determined from the relation
\beq
\label{JointCumulant}
\lim_{t\to\infty} \frac{1}{t}\Esp{\Heate^p \Heato^{p'}}_c =
\left.\frac{\partial^{p+p'} g_{2N}(\lambdae,\lambdao;t)}{\partial \lambdae^p \partial \lambdao^{p'}}\right\vert_{\lambdae=\lambdao=0},
\eeq
where the index $c$ refers to the truncation of the mean value  $\Esp{\Heate^p \Heato^{p'}}$ involved in the definition of the cumulant.
The fact that $g_{2N}$ depends only on the difference $\lambdae-\lambdao$ entails the properties
\beq
\lim_{t\to\infty} \frac{1}{t}\Esp{\Heate^p \Heato^{p'}}_c =(-1)^{p'}\lim_{t\to\infty} \frac{1}{t}\Esp{\Heate^{p+p'}}_c 
\eeq
and, in particular,
\beq
\lim_{t\to\infty} \frac{1}{t}\Esp{\Heato^p}_c =(-1)^{p}\lim_{t\to\infty} \frac{1}{t}\Esp{\Heate^p}_c.
\eeq
These properties are linked to the fact that the interface  energy can take only a finite number of values whereas the cumulants have no upper bounds in the infinite time limit.

For the sake of completeness we point out that, according to \eqref{relthetacosh}, $\theta(\lambdab)$ depends on $\lambdab$ through the function 
$\cosh\left(\lambdab-(\betae-\betao)\gapE/2\right)$; therefore  the scaled generating function satisfies the symmetry $g_{2N}(\lambdao,\lambdae)=g_{2N}(\betao-\lambdao, \betae-\lambdae)$, which is in fact a consequence of the local detailed balance \eqref{LDB}. Since $g_{2N}(\lambdao,\lambdae)$ depends only on the difference $\lambdae-\lambdao$, this entails a symmetry for the scaled generating function for the cumulants of $\Heate$, $g^\text{e}_{2N}(\lambdae)=g_{2N}(0,\lambdae)$, namely the symmetry $g^\text{e}_{2N}(\lambdae)=g^\text{e}_{2N}(\betae-\betao-\lambdae)$. Then the corresponding large deviation function for the time-integrated current ${\cal J}_\text{\mdseries e}=\Heate/t$, which can be obtained as the Legendre-Fenchel transform of $g^\text{e}_{2N}(\lambdae)$, obeys the fluctuation relation $f({\cal J}_\text{\mdseries e})-f(-{\cal J}_\text{\mdseries e})=(\betao-\betae){\cal J}_\text{\mdseries e}$.

\subsection{Cumulants for heat amount $\Heate$}

We give the explicit expressions for the first four cumulants of $\Heate$ per unit of time. They are determined as $\partial^n g_{2N}(0,\lambdae) / \partial\lambda_\text{\mdseries e}^n\vert_{\lambda_\text{\mdseries e}=0}=\gapE^n \times \partial^n g_{2N}(0,\lambdae) / \partial\lambdab^n\vert_{\lambdab=0}$, where $\lambdab=\left(\lambdae-\lambdao\right)\gapE$ and $\gapE=4K$ is the energy gap in the chain energy. The function $\theta(\lambdab)$ defined in \eqref{deftheta} can be rewritten as
\beq
\nuob\nueb \theta(\lambdab)= 2A  [\cosh \lambdab -1] + 2B \sinh \lambdab
\eeq
where the parameters  $A$ and $B$ are those introduced in Ref.\cite{CornuBauer2013} for a model with only two spins, namely
\beq
A=\nuob\nueb (1-\gammao\gammae)
\eeq
\beq
B= \nuob\nueb (\gammao-\gammae).
\eeq
Then, according to \eqref{ExpgL}, the  first four cumulants of $\Heate$ read
\bea
\label{ExpCumulants}
\lim_{t\to\infty} \frac{\Esp{\Heate}}{(\nue+\nuo)t}&=& \frac{N}{2} B \Sig_2 \times \gapE
\\
\nonumber
\lim_{t\to\infty} \frac{\Esp{\Heate^2}_c}{(\nue+\nuo)t}&=& \frac{N}{2} \left[A \Sig_2-B^2\Sig_4\right] \times \gapE^2
\\
\nonumber
\lim_{t\to\infty} \frac{\Esp{\Heate^3}_c}{(\nue+\nuo)t} &=&  \frac{N}{2} B \left[ \Sig_2- 3 A \Sig_4+3B^2 \Sig_6\right] \times \gapE^3
\\
\nonumber
\lim_{t\to\infty} \frac{\Esp{\Heate^4}_c}{(\nue+\nuo)t} &=& \frac{N}{2} \left[A \Sig_2- (3 A^2 + 4B^2) \Sig_4 +18 A B^2 \Sig_6 -15 B^4 \Sig_8 \right] \times \gapE^4,
\eea
where
\beq
\Sig_{2n}(N)=\frac{2}{N}\sum_{k=0}^{(N/2)-1} \sin^{2n} \left(\frac{(2k+1) \pi}{2N} \right).
\eeq
The structure of the cumulants in terms of the coefficients $A$ and $B$ is similar to the structure found for the two-spin model of Ref.\cite{CornuBauer2013} as well as for the ring of  Ref.\cite{CornuHilhorst2017}; indeed, in the three cases the scaled cumulant generating function  depends on $\lambda_\text{\mdseries e}$ only through the same function $\theta(\lambdae \gapE)$.

The coefficients $\Sig_{2n}(N)$ (with $N$ even) can be calculated explicitly. In the special case $N=2$, a direct calculation leads to $\Sig_{2n}(2)=(1/2)^n$. For any $N\geq 2$,
 by extending the summation up to $N$, rewriting $\sin u=[\ed^{\imath u}-\ed^{-\imath u}]/(2\imath)$, using the binomial formula and an identity  similar to \eqref{SumGeom}, we get that
\beq
\label{Wallis1}
\textrm{if $n<N$}\quad \Sig_{2n}(N)=\left(\frac{1}{2} \right)^{2n} \frac{[2n]!}{(n!)^2}=W_{2n},
\eeq
where $W_{2n}$ denotes the normalized Wallis integral, $W_{2n}=(2/\pi) \int_0^{\pi/2} (\sin q)^{2n} \dd q$. The first four values of the latter integrals are
\beq
\label{valueWallis}
W_2=\frac{1}{2}, \qquad  W_4=\frac{3}{8}, \qquad W_6=\frac{5}{16},  \qquad \textrm{and} \quad 
W_8=\frac{35}{128}.
\eeq
For  $n$ larger than $N$,  a finite-size correction arises. For instance,
\beq
\textrm{if $N\leq n< 2N$}\quad \Sig_{2n}(N)=W_{2n}\left[ 1-2 \frac{[n!]^2}{(n-N)! (n+N)! }\right].
\eeq

From the study of the $S_{2n}(N)$ we get that  all cumulants of order $n$ smaller than  $N$, the number of sites connected to a given thermostat, are strictly proportional to the size $L=2N$ of the chain. Finite-size corrections appear only in cumulants of order $n\geq N$. 

In particular, $S_2(N)=1/2$ for all $N$, by virtue of  \eqref{Wallis1}-\eqref{valueWallis} valid for all $N\geq 2$ with $N$ even. 
Therefore  the first cumulant $\lim_{t\to +\infty}\Esp{\Heate}/t$ given in  \eqref{ExpCumulants} does coincide with the expression \eqref{valueJe} of the mean  global instantaneous current $\Espst{J_\text{\mdseries e}}$ in the stationary state. We have already pointed out that the latter mean global current  contains no finite-size corrections. If $N>4$ the first four cumulants are given by \eqref{ExpCumulants} where the $S_{2n}(N)$ are to be replaced by the corresponding $W_{2n}$ given in \eqref{valueWallis}.

Eventually the dependences upon the thermodynamic parameters, $\gammao$ and $\gammae$, and  the kinetic parameters, $\nuo$ and $\nue$, arise only through the coefficients $A$ and $B$. This is in contrast with what happens for another   Ising chain model, where both thermostats act on every spin \cite{CornuHilhorst2017}: for this model the dependence upon the combination $\gamma$ of the thermodynamic and kinetic parameters defined in \eqref{defgamma} also arises in the coefficients $\Sigma_n(N,\gamma)$ which replace the coefficients $\Sig_n(N)$ in the expressions   \eqref{ExpCumulants} for the cumulants.

\clearpage
\subsection{Various physical regimes}

According to the last remark of the previous section, the discussion of the various physical regimes is the same as that performed in the case of the two-spin model of Ref.\cite{CornuBauer2013}.

At equilibrium $\gammao=\gammae$ and, according to \eqref{deftheta}, $\theta(\lambdab)=2(1-\gammae^2) (\cosh \lambdab -1)$. Therefore only cumulants of even order do not vanish. According to \eqref{ExpCumulants} the first two cumulants of even order read
\bea
\label{ExpCumulantsEq}
\lim_{t\to\infty} \frac{\Esp{\Heate^2}_c}{(\nue+\nuo)t}&=& \frac{1}{4} \nuo\nue \left( 1-\gammae^2\right) \times N \gapE^2
\\
\nonumber
\lim_{t\to\infty} \frac{\Esp{\Heate^4}_c}{(\nue+\nuo)t} &=&  \frac{1}{4} \nuo\nue \left( 1-\gammae^2\right) \left[1-\frac{9}{4}\nuo\nue \left( 1-\gammae^2\right)  \right]\times N\gapE^4.
\eea
The probability distribution of $\Heate$ is not a Gaussian, since all cumulants of even order have non-vanishing values.

\medskip

When a thermostat has a kinetic parameter far larger than the other one,  the scaled generating function becomes proportional to $\theta(\lambdab)$, 
\beq
g_{2N}(\lambdab)=\frac{1}{8} N \nus\theta(\lambdab),
\eeq
where $\nus$ is the kinetic parameter of the slower thermostat. As a consequence $g_{2N}(\lambdab)$
coincides with the scaled generating function of a continuous-time random walk, because $\theta(\lambdab)$ can be rewritten as
\beq
\theta(\lambdab)= 2 \left[ p_+ \ed^{\lambdab} +p_- \ed^{-\lambdab} -(p_++p_-)\right]
\eeq
with the probabilities $p_+=(1+\gammao)(1-\gammae)/2$ and $p_-=(1-\gammao)(1+\gammae)/2$. As a consequence all cumulants of even  (odd) order are equal to the same value when they are measured in unit of $\gapE$: for all $p\geq 1$
\bea
\lim_{t\to\infty} \frac{\Esp{\Heate^{2p-1}}_c}{ t \, \gapE^{2p-1}} &=& \frac{1}{4} \nus (\gammao-\gammae) \,  N
\\
\nonumber
\lim_{t\to\infty} \frac{\Esp{\Heate^{2p}}_c}{t \,  \gapE^{2p}} &=& \frac{1}{4} \nus (1-\gammao\gammae) \, N.
\eea
In the same kinetic regime, if  one thermostat is at zero temperature, for instance $\gammao=1$, then
the scaled generating function  coincides with that of a continuous-time  Poisson process, because
\beq
\theta(\lambdab)= 2 (1-\gammae) \left[ \ed^{\lambdab}-1\right].
\eeq
As a consequence all cumulants  of $\Heate$ in unit of $\gapE$ are equal to the same value: for all $p\geq 1$
\beq
\lim_{t\to\infty} \frac{\Esp{\Heate^{p}}_c}{t \,  \gapE^p} = \frac{1}{4} \nus (1-\gammae)  \, N.
\eeq

\section{Conclusion}

In the present paper we have investigated the heat currents in an Ising spin ring where alternating spins are coupled to two macroscopic bodies at different temperatures and with different kinetic parameters. The stationary mean values  of the global two-spin correlations at any distance have been calculated. The dependence upon the kinetic parameters arises only through the linear combination $\gamma$ defined in \eqref{defgamma}\footnote{When the kinetic parameters are set equal,  our result are compatible with those of Ref.\cite{SchmittmannSchmuser2002}.}. The finite-size corrections in the global two-spin correlations disappear in the mean instantaneous global  heat current coming out of one thermostat.

The scaled generating function of the joint cumulants per unit of time for the heat amounts exchanged with the two thermostats over a long time have been calculated exactly. At leading order in the ring size they prove to be  proportional to the ring size, as it is the case for the model where two thermostats act on the same site \cite{CornuHilhorst2017}.
 Moreover, if the order of the cumulant is lower than the number of  spins connected to one thermostat, the finite-size corrections again disappear exactly, and the cumulant is strictly proportional to the ring size.

 We notice that the proportionality  to the ring size at leading order in the size has already been observed for the cumulants of two other kinds of cumulative quantities  when the system is homogeneous (only one temperature and one kinetic parameter) and seen as a Simple Exclusion Process with pair creation and annihilation \cite{PopkovSchutz2011}: the two cumulative quantities are the difference between the numbers of domain wall jumps  in the clockwise and anticlockwise directions respectively, and  the number of pair annihilations. This is in contrast with the case of the purely diffusive Simple Exclusion Process on a ring, where the cumulants for the difference between the numbers of jumps in the two directions and the cumulants for  the total number of jumps are proportional to powers of the ring size which increase with the order of the cumulants \cite{AppertDerridaETAL2008}.

An interesting open problem is the calculation of the heat cumulants in another solvable model for thermal contact : two joined semi-infinite Ising chains 
coupled to thermostats at two different temperatures \cite{LavrentovichZia2010,Lavrentovich2012}. Then the mean global current which flows from one thermostat to the other through the junction between the two half-chains is obtained by summing the  mean  currents received by all spins in a semi-infinite Ising chain. The intrinsic inhomogeneity of these currents
would have to be dealt with by specific methods.

\appendix
\section{Two-spin correlations}
\label{secTwospincorrelations}

By analogy with the homogeneous case \cite{Glauber1963} the evolution equation for the two-spin correlations reads
\beq
\frac{\dd \Esp{s_j s_k}}{\dd t} = - 2 \Esp{s_js_k\left[ w(s_j \to -s_j) +w(s_k \to -s_k)  \right] }.
\eeq
Inserting the expression \eqref{deftauxtrans} for the transition rates we get
\beq
\label{EqEvolutionCorr}
\frac{\dd\Esp{s_j s_k}}{\dd t} = - (\nu_j +\nu_k)\Esp{s_js_k} +
\tfrac{1}{2}\nu_j\gamma_j \left[ \Esp{s_{j-1}s_k} +\Esp{s_{j+1}s_k} \right]
+ \tfrac{1}{2}\nu_k\gamma_k \left[ \Esp{s_js_{k-1}} +\Esp{s_js_{k+1}} \right].
\eeq

The latter equations imply that $D^\text{ee}_2$, $D^\text{oo}_2$, $D^\text{oe}_1$ and $D^\text{eo}_1$ involved in the expressions  \eqref{ExpressionJe} and \eqref{ExpressionJo} for the  mean global currents $\Espst{J_\text{\mdseries e}}$ and $\Espst{J_\text{\mdseries o}}$ are to be determined from a hierarchy of equations for the two-spin quantities 
\beq
\label{DefDoo}
D^\text{oo}_{2p}=\frac{1}{N}\sum_{n=1}^N \Espst{s_{2n-1} s_{2n-1+2p}},
\eeq
\beq
\label{DefDoe}
D^\text{oe}_{2p+1}=\frac{1}{N}\sum_{k=1}^N \Espst{s_{2n-1} s_{2n+2p}}
\eeq
with an analogous definition for $D^\text{eo}_{2p+1}$, and 
\beq
\label{DefDee}
D^\text{ee}_{2p}=\frac{1}{N}\sum_{n=1}^N \Espst{s_{2n} s_{2n+2p}}.
\eeq
 We notice that if the initial probability distribution for the spin configurations is translationally invariant, this property is conserved by the evolution under the transition rates and the stationary two-spin correlation $\Espst{s_k s_{k+p}}$ depends only on the difference $p$ between the site indices; then it is equal to one of the $D$'s defined  in \eqref{DefDoo},
\eqref{DefDoe} and \eqref{DefDee}. 
From the latter definitions we get the boundary conditions
\beq
\label{BC1}
D^\text{ee}_0=1   \quad\textrm{and}\quad D^\text{oo}_0=1.
\eeq

From the evolution equation for the spin correlations \eqref{EqEvolutionCorr} we get that, in the stationary state where the $D$'s are defined, for $2\leq 2p \leq 2(N-1)$
\bea
\label{EqDee}
0&=&-4\nueb D^\text{ee}_{2p} + \nueb \gammae \left[ D^\text{oe}_{2p+1} + D^\text{oe}_{2p-1} +D^\text{eo}_{2p-1}  +D^\text{eo}_{2p+1} \right]
\\
\label{EqDoo}
0&=&-4\nuob D^\text{oo}_{2p} + \nuob \gammao \left[ D^\text{eo}_{2p+1} + D^\text{eo}_{2p-1} +D^\text{oe}_{2p-1}  +D^\text{oe}_{2p+1} \right],
\eea
while for $1\leq 2p+1 \leq 2N-1$
\bea
\label{EqDoe}
0&=&-2 D^\text{oe}_{2p+1} + \nuob \gammao \left[ D^\text{ee}_{2p+2} + D^\text{ee}_{2p}\right] + 
\nueb \gammae \left[D^\text{oo}_{2p}  +D^\text{oo}_{2p+2} \right]
\\
\label{EqDeo}
0&=&-2 D^\text{eo}_{2p+1} + \nuob \gammao \left[ D^\text{ee}_{2p+2} + D^\text{ee}_{2p}\right] + 
\nueb \gammae \left[D^\text{oo}_{2p}  +D^\text{oo}_{2p+2} \right].
\eea
Comparison of   equations \eqref{EqDee} and \eqref{EqDoo} leads to the relation valid for $2\leq 2p\leq 2(N-1)$
\beq
\label{relDooDee}
D^\text{oo}_{2p}=\frac{\gammao}{\gammae}D^\text{ee}_{2p},
\eeq
while comparison of equations \eqref{EqDoe} and \eqref{EqDeo} leads to the relation valid for $1\leq 2p+1 \leq 2N-1$
\beq
\label{relDoeDeo}
D^\text{eo}_{2p+1}=D^\text{oe}_{2p+1}.
\eeq
By taking into account these relations in \eqref{EqDee} and \eqref{EqDoe} we have to solve the coupled equations for the $D^\text{ee}$'s and the $D^\text{oe}$'s 
\bea
\label{Recursive}
-2 D^\text{ee}_{2p} + \gammae\left[ D^\text{oe}_{2p-1} +D^\text{oe}_{2p+1} \right] &=& 0
 \quad\textrm{for}\quad 2 \leq 2p \leq 2(N-1)
\\
\nonumber
-2 D^\text{oe}_{2p+1} + \gammao\left[ D^\text{ee}_{2p} +D^\text{ee}_{2p+2} \right] &=& 0
 \quad\textrm{for}\quad 2 \leq 2p \leq 2(N-2),
\eea
where the second equation is to be supplemented by  the extra boundary conditions for $p=0$ and $p=N-1$ respectively. The latter conditions  are derived from \eqref{EqDoe} and \eqref{BC1},
\bea
\label{BC2}
-2 D^\text{oe}_1 + \gammao D^\text{ee}_2 + \gamma &=& 0
\\
\nonumber
-2 D^\text{oe}_{2N-1} + \gammao D^\text{ee}_{2(N-1)} + \gamma &=& 0,
\eea
where $\gamma$ is defined in \eqref{defgamma}.
The  equations \eqref{Recursive} allow to determine recursively $D^\text{oe}_3, D^\text{ee}_4, \ldots, D^\text{oe}_{2N-1} $ from a given set $(D^\text{oe}_1, D^\text{ee}_2)$, and then the boundary conditions \eqref{BC2} determine the values of $D^\text{oe}_1$ and $D^\text{ee}_2$. 

The  recursive equations \eqref{Recursive} are linear and their generic solution, which depends on the two parameters $D^\text{oe}_1$ and $D^\text{ee}_2$, can be looked for as  linear combinations of two linearly independent solutions. Because of the  invariance of these equations under the translation over two sites, one can look for independent solutions which are also eigenfunctions of the translation operator on each sublattice, namely solutions of the form $f^\text{ee}_{2p+2}=\eta f^\text{ee}_{2p}$ and  $f^\text{oe}_{2p+3}=\eta f^\text{oe}_{2p+1}$. These solutions can be written as
\bea
\label{TranslationInvariant}
f^\text{ee}_{2p} &=& \eta^{p-1} a
\\
\nonumber
f^\text{oe}_{2p+1} &=&\eta^p b.
\eea
By inserting the latter expressions into the recursive equations \eqref{Recursive} one gets two coupled linear equations for $a$ and $b$. The latter do not vanish if $\eta$ is equal to one of the two values 
\beq
\label{Valueetapm}
\eta_\pm= \frac{2-\gammao\gammae \pm 2 \sqrt{1-\gammao\gammae}}{\gammao\gammae}
=\left[ \frac{1 \pm \sqrt{1-\gammao\gammae}}{\sqrt{\gammao\gammae}} \right]^2.
\eeq
The model is defined for $0<\gammao<1$ and $0<\gammae<1$, so that $\eta_+\neq \eta_-$.
Then in the two solutions of \eqref{TranslationInvariant} with $\eta_+$ and $\eta_-$ respectively, $b_\pm= \tfrac{1}{2} \gammao [1+\eta_\pm^{-1}] a_\pm$. Then 
the generic solution of \eqref{Recursive} reads
\bea
D^\text{ee}_{2p} &=&  a_+\eta_+^{p-1}  + a_- \eta_-^{p-1} 
\\
\nonumber
D^\text{oo}_{2p} &=&   \tfrac{1}{2}\gammao (1+\eta_+^{-1}) a_+\eta_+^p  + \tfrac{1}{2}\gammao (1+\eta_-^{-1})
 a_- \eta_-^p,
 \eea
 In fact the solution can be written only in terms of $\eta_-$ (with $0< \eta <1$) by using the relation $\eta_+\eta_-=1$.
Then the boundary conditions \eqref{BC2} determine the values of $a_+$ and $a_-$. After straightforward calculations and use of the relation
\beq
\label{RelRoots}
\sqrt{\eta_-}+ \frac{1}{\sqrt{\eta_-}}=\frac{2}{\sqrt{\gammao\gammae}},
\eeq
we get that for $2\leq 2p\leq 2(N-1)$
\beq
\label{valueDee}
D^\text{ee}_{2p}=\frac{\gamma}{\gammao} \frac{1}{1+\eta_-^N} 
\left[ \eta_-^{p} + \eta_-^{N- p}  \right],
\eeq
and for $1\leq 2p+1 \leq 2N-1$
\beq
\label{valueDoe}
D^\text{oe}_{2p+1}=\frac{\gamma}{\sqrt{\gammao\gammae}} \frac{1}{1+\eta_-^N} 
\left[ \eta_-^{p+1/2} + \eta_-^{N- p-1/2}  \right].
\eeq
The latter formulae are compatible with the expressions of the stationary two-spin correlations determined in Ref.\cite{SchmittmannSchmuser2002} in the case where $\nuo=\nue$.

\bibliographystyle{unsrt}

\end{document}